# Towards fully *ab initio* simulation of atmospheric aerosol nucleation

Shuai Jiang[1], Yi-Rong Liu[1], Teng Huang[3], Ya-Juan Feng[1], Chun-Yu Wang[1], Zhong-Quan Wang[3], Wei Huang[1, 2, 3*]

[1]*School of Information Science and Technology, University of Science and Technology of China, Hefei, Anhui 230026, China*
[2]*Center for Excellent in Urban Atmospheric Environment, Institute of Urban Environment, Chinese Academy of Sciences, Xiamen, Fujian 361021, China*
[3]*Laboratory of Atmospheric Physico-Chemistry, Anhui Institute of Optics & Fine Mechanics, Chinese Academy of Sciences, Hefei, Anhui 230031, China*
[*]email: huangwei6@ustc.edu.cn

**Atmospheric aerosol nucleation contributes to more than half of cloud condensation nuclei globally[1]. The emissions, properties and concentrations of atmospheric aerosols or aerosol precursors could respond significantly to climate change[2]. Despite the importance for climate, the detailed nucleation mechanisms are still poorly understood. The ultimate goal of theoretical understanding aerosol nucleation is to simulate nucleation in ambient condition, hindered by lack of accurate reactive force field[3]. Here we propose the reactive force field for nucleation systems with good size scalability based on deep neural network. The huge computational costs from direct molecular dynamics in ambient conditions are surmounted by bridging the simulation in the limited box with cluster kinetics, facilitating the aerosol nucleation simulation to be fully *ab initio*. We found that the acid-base formation rates previously based on hard sphere collision rate constants tend to be underestimated up to several times. These findings show that the widely recognized acid-base nucleation observed in the CLOUD (Cosmics Leaving OUtdoor Droplets) chamber experiments[4-7], pristine[8] and polluted[9-11] environments should be revisited to considering the contribution of collision enhancement. Besides, the framework here is transferable to other nucleation systems[12], potentially boosting the nucleation parameterizations accuracy generally to effectively advance the climate model predictions reliability.**

Theoretical understanding of nucleation mechanism largely relies on classical nucleation theory (CNT)[13] originally proposed in 1935, which gives a general mind map for nucleation thermodynamics and kinetics[3] even though the capillary assumption is extensively criticized[14]. Emergence[15] in 2011 and then broadly employed[5,9,16-19] theoretical model, Atmospheric Cluster Dynamics Code (ACDC), surmounts the drawbacks of CNT through coupled quantum chemical thermodynamics[20] with the birth-death equations[3]. Within ACDC, collision rate constants and evaporation rates are the two most critical parameters, determining the accuracy to predict macro parameters like cluster concentrations and formation rates which can be directly compared with experiments[5]. Evaporation rates, derived from detailed balance and *ab initio* thermodynamics, can be very accurate with sophisticated quantum chemical calculations[21]. However, collision rate constants, derived from simple hard sphere model, are still very rough, which accuracy is far from the *ab initio* based evaporation rates'. Moreover, the accuracy of collision rate constants is extremely important especially for collision-controlled systems like sulfuric acid-dimethylamine system as evaporation rates are close to zero[6]. Pioneering work[22] investigated the collisions between sulfuric acid monomers, however, the force field utilized there lacks reactivity and the computational costs of extending the method on more collisions among molecules and/or clusters are enormous. So highly accurate and computationally cheap reactive force field for flexible nucleation clusters is urgently needed to simulate nucleation processes with fully *ab initio*.

Here we propose a general framework to potentially boost the aerosol nucleation simulation toward fully *ab initio*. In the framework, comprehensive data sets are firstly prepared though passive learning coupled with active learning techniques. Then deep neural network based force field (DNN-FF) is trained so that robust nucleation molecular dynamics (MD) simulations can be performed to derive the collision rate constants based on Poisson distribution. Then static quantum chemical thermodynamics based evaporation rates are coupled with DNN-FF based MD derived collision rate constants into cluster dynamics model to provide *ab initio* kinetics for simulating atmospheric aerosol nucleation.

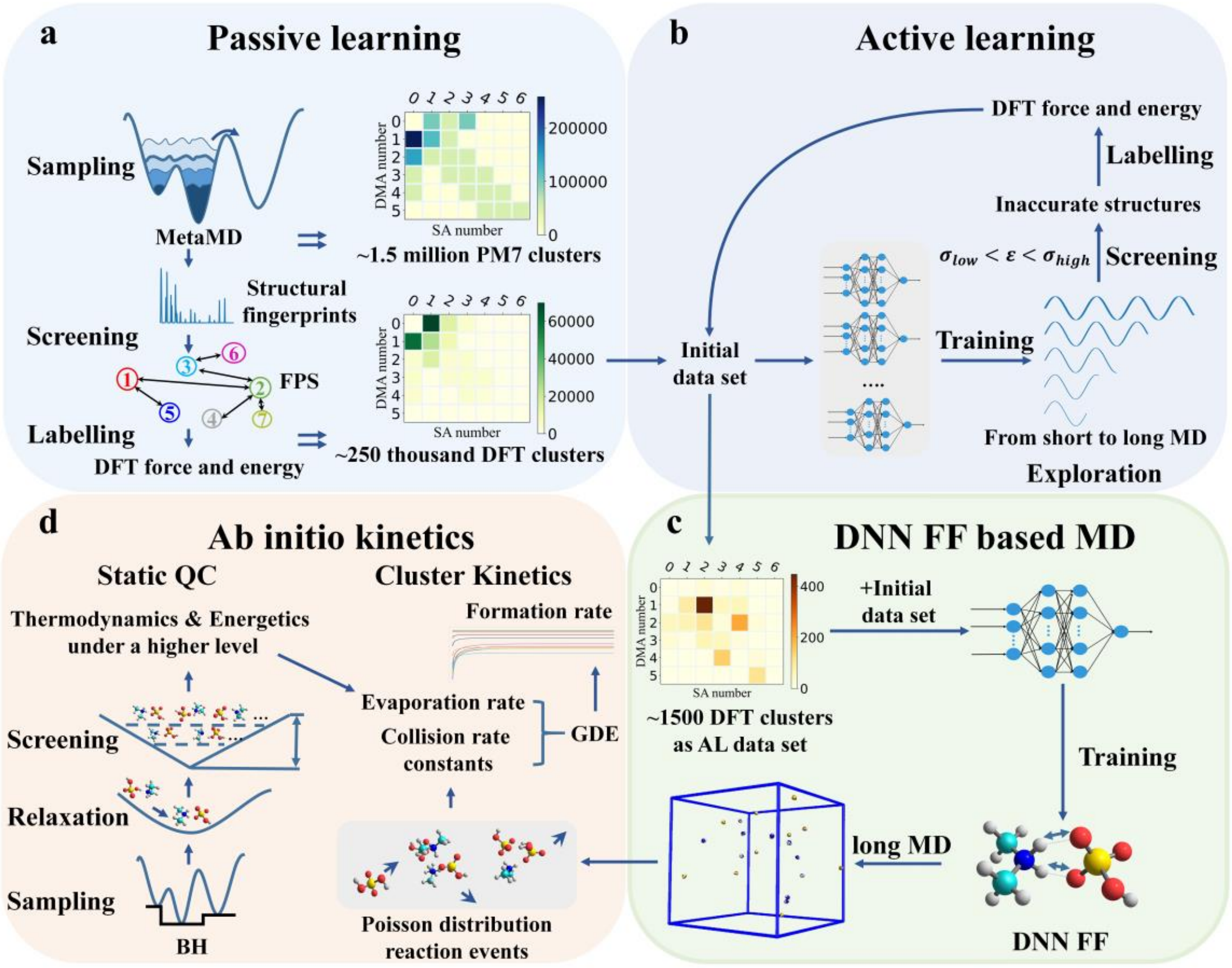

**Figure 1 | A general framework towards fully *ab initio* simulation of atmospheric aerosol nucleation. a** and **b** shows passive learning and active learning techniques to prepare data set for DNN respectively. **c**, DNN-FF driven MD. **d**, Cluster kinetics simulation based on MD derived collision rate constants and static quantum chemistry (QC) calculation derived evaporation rates.

The general framework towards fully *ab initio* simulation of aerosol nucleation is shown in Fig. 1. The details in each module can be found in method section, so here the major points about significance and correlation for each module are given. Here passive learning is no doubt very important as the provided data set occupies most of the final data set (99.37 % in Extended Data Table 3). However, there are two main shortcomings for passive learning. Firstly, cluster composition in each passive learning is set beforehand, so traversing all compositions for limited cluster size would undoubtedly produce redundant data set. Secondly, the sampling and subsequent screening structure number is empirical. Comparatively, active learning is very critical despite of the very small proportion of final data set (0.63 % in Extended Data Table 3). To be noted, tests show purely based on passive learning provided data set cannot provide a robust DNN model as unphysical behaviors are observed, that's why active learning is introduced afterwards. The cluster size sampled by passive learning is based on the experimental experience that acid-base clusters are mostly stable when the difference between acid number and base number is less than or equal to one[5]. Active learning not only supplements the structures for passive learning sampled size but also points to the cluster compositions with high evaporation rates, e.g. 4DMA as we can see from active learning data set in Fig. 1c, which can normally be ignored through passive learning.

Currently the coupling between machine learning (ML) and chemistry is on the rise, so training a DNN model with good performance on train and test data sets are becoming more and more routine[23-31]. However, training a DNN model with good size scalability and applicable on reactive MD simulation, especially for flexible molecules in this case is still very challenging[32]. Robust MD simulations here prove the high quality of data set given by passive learning and active learning as well as the excellent performance of descriptors combined with neural network framework and parameters.

Due to the interpolate nature of DNN, size scalability is still limited for extending only several molecules in this case. However, considering the rare event nature[33] of aerosol nucleation, direct MD for ambient or laboratory aerosol nucleation with high concentration is very expensive[34], challenging the application value of DNN trained model. We bridge the gap between micro parameters and macro ones through embedding MD derived rate constants based on Poisson distribution into cluster kinetics model. Those DNN-FF based MD derived constants coupling with static QC derived evaporation rates effectively boost the aerosol simulation towards fully *ab initio*.

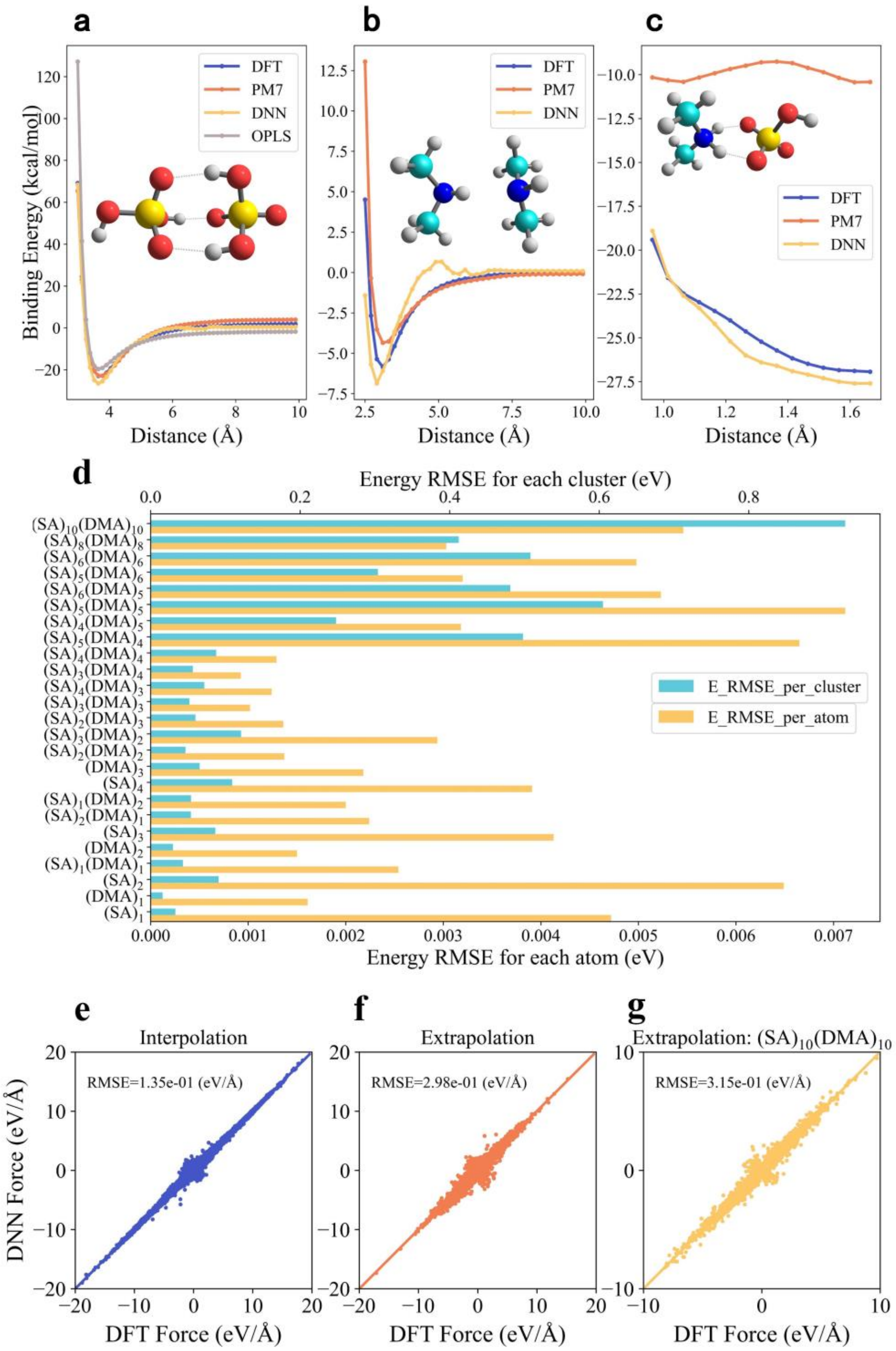

**Figure 2 | DNN force field benchmark. a, b** and **c** give the dimer detachment curves for sulfuric acid-sulfuric acid (SA-SA), dimethylamine-dimethylamine (DMA-DMA) and sulfuric acid-dimethylamine (SA-DMA), respectively. **d,** Energy root mean squared error (RMSE, in eV) for each cluster (in blue) or each atom within cluster (in yellow). **e,** Force interpolation (eV/Å). **f,** Force extrapolation (eV/Å). **g.** Force extrapolation (eV/Å) for $(SA)_{10}(DMA)_{10}$.

The dimer detachment curves in Fig. 2a-c provide the basic picture about the performance of various force fields. For root mean squared error (RMSE) of binding energy (Extended Data Table 1), DNN is clearly superior than PM7 semi-empirical method[35] and optimized potentials for liquid simulation (OPLS)[36]. Notably, DNN performs largely better than OPLS for sulfuric acid (SA) dimer and PM7 for sulfuric acid-dimethylamine (SA-DMA) dimer. For smoothness, PM7 and OPLS is smoother than DNN generally, it is probably due to the absence of electrostatic interactions beyond the cutoff[37], however, the novelty and validity of the proposed framework is not affected. Despite of the shortcoming, nanosecond MD simulation based on the DNN-FF is still working without observing unphysical behaviors like molecules dissociating into atoms. The cluster size resolved energy RMSE for each atom (Fig. 2d) shows high accuracy with the largest value lowing than $7 \times 10^{-2}$ eV. The maximum cluster size within the training set is $(SA)_5(DMA)_6$. To be noted, $(SA)_m(DMA)_n$ represents a cluster with m number of sulfuric acid molecules and n number of dimethylamine molecules. The size extension to $(SA)_{10}(DMA)_{10}$ gives quite good accuracy with $2.98 \times 10^{-1}$ eV/Å for all cluster sizes considered here (Fig. 2f) and $3.15 \times 10^{-1}$ eV/Å for $(SA)_{10}(DMA)_{10}$ alone (Fig. 2g). In a word, DNN model's superior performance in energy and force descriptions in addition to its distinguished size scalability lay a solid foundation for robust nucleation MD simulations.

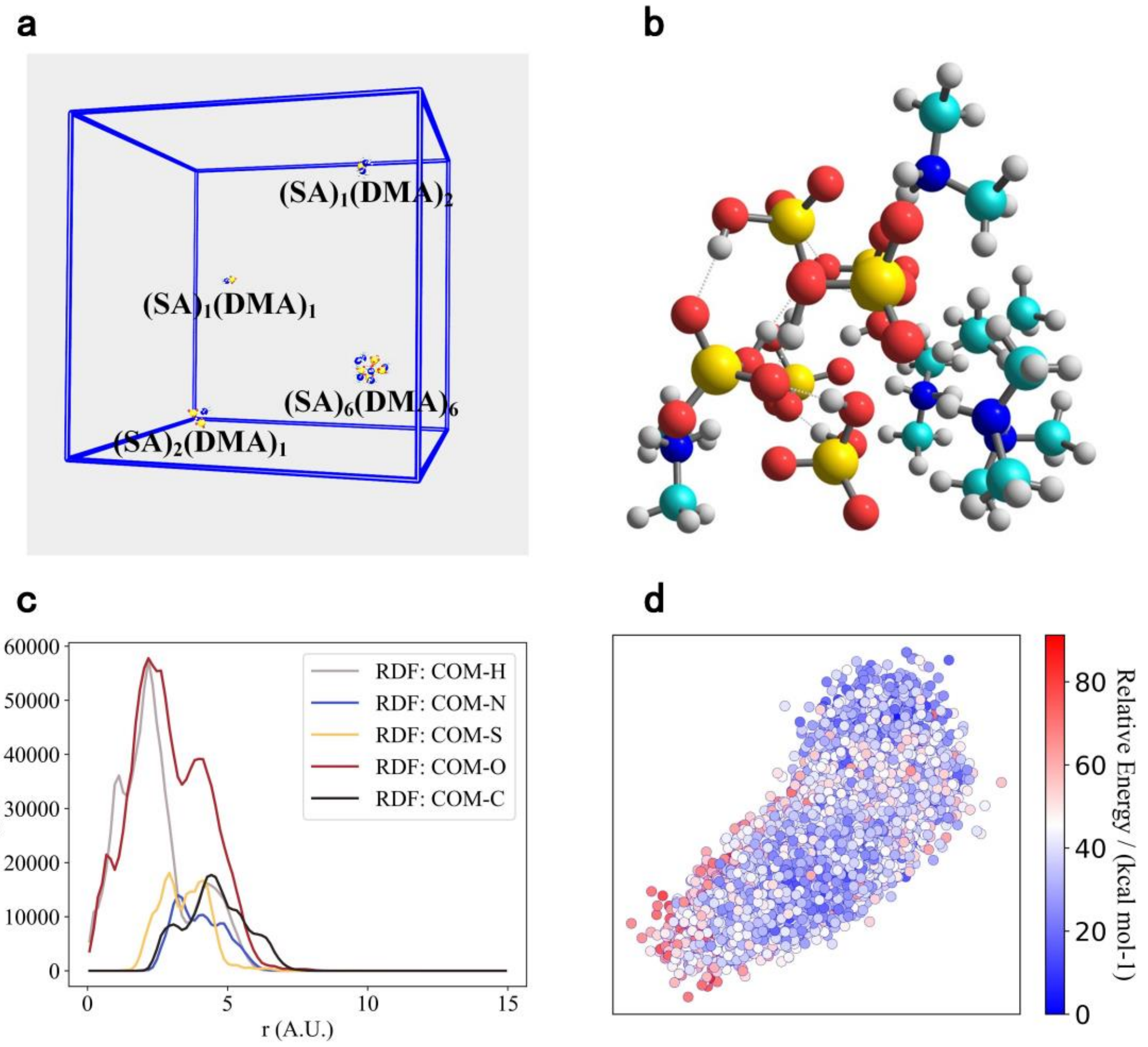

**Figure 3 | Structural distribution for $(SA)_6(DMA)_6$ isomers derived from DNN-FF based MD. a** shows the snapshots of MD for at 1 ns. The cyan, white, red, blue and yellow circles represent C, H, O, N and S atoms respectively. The N and S atoms radii are enlarged three times for clarity. **b,** The most stable isomer in trajectory. **c,** Radial distribution function (RDF) between cluster center of mass (COM) and the five elements. **d,** Kernel principal component analysis (KPCA)[38] maps of isomers using a global Smooth Overlap of Atomic Positions (SOAP)[39] kernel.

With robust size scalability of DNN model, nanosecond scale MD simulation for cluster collision and evaporation can be made, whose representative snapshot is shown in Fig. 3a. Furthermore, the isolated cluster can be singled out for gaining insights about their structural evolution. Here $(SA)_6(DMA)_6$ is chosen since it is the largest cluster with the same number of acids and bases observed in DNN-FF based MD, also very close to the lowest experimentally detectable size (~1.7 nm)[5]. Surprisingly, three instead of all dimethylamine molecules are pronated initially (This is why only nine H-N distances are plotted in Extended Data Figure 5). For the most stable isomer (Fig. 3b), the sulfuric acid molecules are bonded with each other through hydrogen bonds, formed a shell with the cluster center of mass (COM) inside. The phenomena that the hydrogen bonded sulfuric acid molecules forms the first shell is general across all the $(SA)_6(DMA)_6$ isomers as shown in the radial distribution function (RDF) from Fig. 3c. The more inner position of COM-S and COM-O than that of COM-N highlights the feature of the first shell composing of sulfuric acid molecules. The structural similarity can also be seen through the closely connected points in the energy basin (Fig. 3d). Besides, here $(SA)_6(DMA)_6$ cluster is from the collision between $(SA)_5(DMA)_5$ and $(SA)_1(DMA)_1$ and the high energy isomers during colliding and subsequently rearrangements can also be seen in Fig. 3d (semitransparent red points in the lower left corner). Interestingly, one proton transfer event among dimethylamine molecules occurs at around 963 ps (Extended Data Figure 5). Comparatively, tons of proton transfer events occur among sulfuric acid molecules, making the proton number within one sulfuric acid molecules jumping from zero to three (Extended Data Figure 6). Notably, although sulfuric acid without protons or with three protons is rarely seen, we see several events whose duration is more than 100 fs. From above analysis, essential structural insights can be obtained through collecting the clusters with the same composition from MD trajectory.

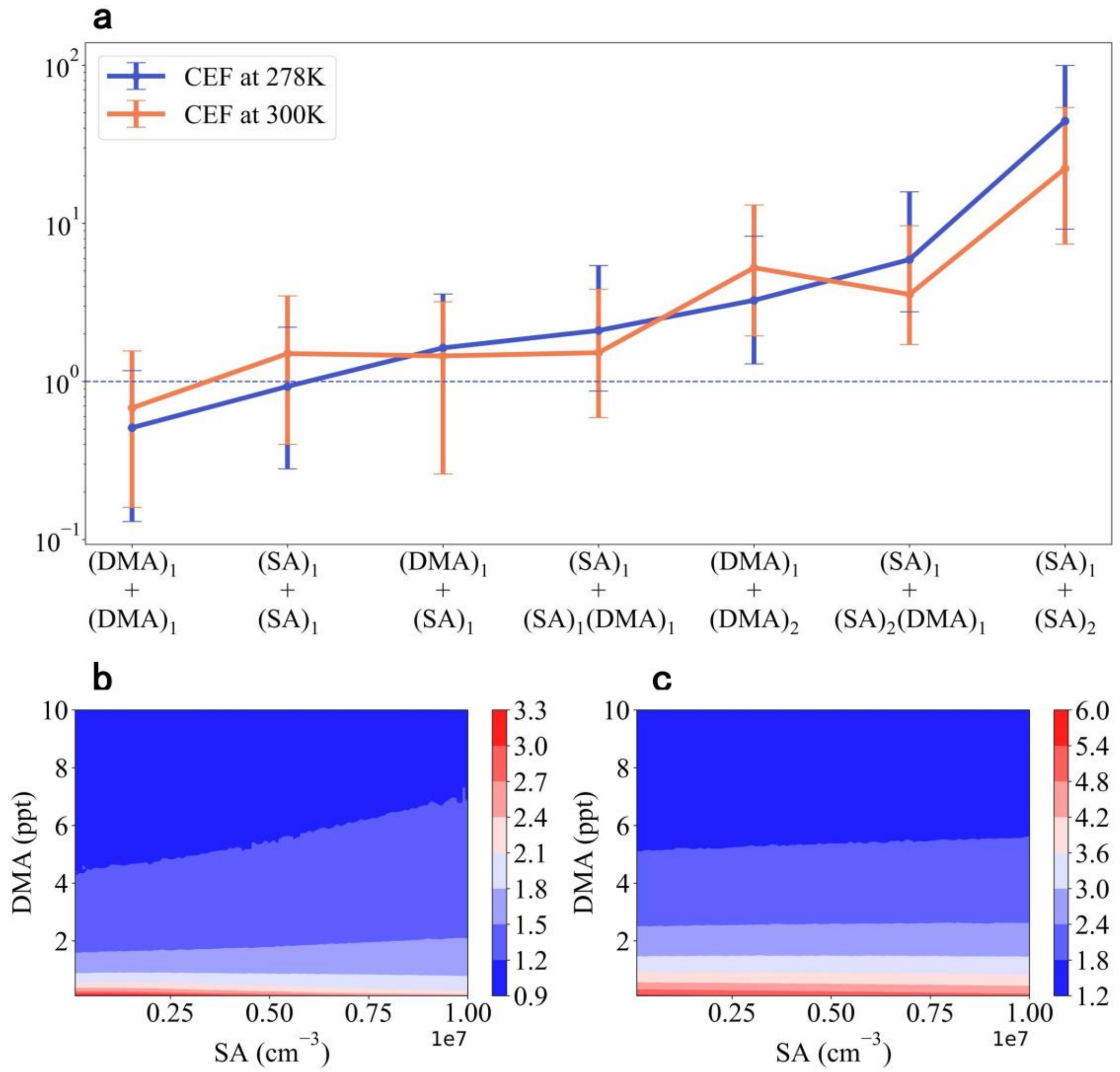

**Figure 4 | Collision and Formation Rate Enhancement Factor (CEF and FREF). a**, The collision and temperature dependence of CEF. Here only reactions both appearing at 278 K and 300 K are shown, the full reaction list can be found in Extended Data Table 2. **b** and **c** give the FREF for representative clean (T=278 K, CS=2.6 × $10^{-3}$ $s^{-1}$) and polluted environment (T=278 K, CS=2.7 × $10^{-2}$ $s^{-1}$) respectively.

DNN-FF driven MD simulation of molecular clusters collisions and evaporations follow Poisson distribution[40], so that the so-called Collision Enhancement Factor (CEF) can be derived. It is the quotient of MD derived collision rate constants divided by hard spheres collision rate constants. CEF is typically more than one due to the long-range intermolecular forces[22]. However, for the collision between dimethylamine monomer, CEF is below one mainly because of the intermolecular repulsion of dimethylamine molecules (Fig. 4a). To be noted, considering that long-range interaction in DNN is limited within the cutoff, the CEF here is certainly underestimated. Despite this, this shortcoming doesn't affect the validity and application value of the framework. The temperature dependence of CEF is not strong and not clearly distinguished comparing that from 278 K with 300 K. Moreover, the composition dependence of CEF is also not clear, but for most collisions, CEF is more than one. Replacing hard spheres collision rate constants with MD derived ones in cluster kinetics model paves the way for fully *ab initio* simulation of aerosol nucleation. The derived formation rates divided by those based hard spheres collision rate constants give FREF (Fig. 4b and 4c). In representative clean (Fig.4b) and polluted (Fig. 4c) environment, FREF has strong negative correlation with DMA concentration and weak negative correlation with SA concentration. Under the same nucleation precursor concentration, FREF in polluted environments is larger than that in clean environment. Considering typically less precursor concentrations in clean environment than those in polluted environment, which one case has the larger FREF depends on the specific precursor concentrations. Generally speaking, here FREF ranges from one to six or so, however, this is the lower bound values as still there are quite a number of collision rate constants needing to be replaced and the underestimation of CEF due to lack of electrostatic interaction beyond the cutoff in DNN-FF. So the larger formation rates than expected before challenges the idea that SA-DMA nucleation is collision-limited with zero evaporation rates[7], providing the alternative scenario that collision rate constants are underestimated while evaporation rates are low but not zero. Further studies about this are definitely needed when fully *ab initio* kinetics are available in the future.

Further directions lie in how to produce compact data set first, probably heavily relying on active learning. Another aspect is to improve the DNN model accuracy, possibly through transfer learning[41] or Δ-learning[42]. For nucleation applications, more diverse box size and initial monomer spatial distribution MD simulations need to be conducted to lower the uncertainties[40] to estimate collision rate constants. Besides, long-range interaction beyond the cutoff size (6 Å here) should be carefully considered in addition that larger box size simulation with more molecules inside is needed to include more various collisions, so that the simulation could be fully *ab initio*. Afterwards, the lots of *ab initio* derived collision rates could tentatively be predicted purely by molecular physical chemistry properties to save simulation time costs.

The complex and various nucleation precursors in ambient environment, especially in polluted environment, call for new theoretical methods in addition to static QC to unravel the complication mechanisms. We believe the framework proposed here that the introducing of DNN-FF and the bridging between MD derived rate constants with cluster kinetics paves towards the fully *ab initio* simulation of atmosphere aerosol nucleation. The high accurate formation rate derived here can be further parametrized into climate model to improve the climate prediction in a global and local scale.

## Method

**Passive learning.** Instead of sampled by Basin-Hopping[43] that is already widely applied in atmospheric non-covalent interaction clusters[44], the nucleation clusters PES is sampled by metadynamics (MetaMD)[45] due to its ability to sample high energy isomers, to prepare the initial data set for further active learning. The bump perturbation can be calculated as

$$V_{bump}(\vec{R}) = \sum_{\alpha} \lambda e^{-\sum_{ij} \frac{D_{ij}(\vec{R}) - D_{ij}^{\alpha}}{2\sigma^2}} \tag{1}$$

Here $\alpha$ sums over snapshots of geometries where a "bump" has been placed at a particular value of the collective variable (CV), and $D_{ij}$ is the (possibly sparse) atomic distance matrix at a given point during the trajectory. The MD temperature, bump width $\sigma$ and bump height $\lambda$ is set to be 600 K, 2.0 and 1.0 respectively. Clusters whose size is within the range of $(SA)_m(DMA)_n$ (m=0-4, n=0-4) is sampled with MetaMD coupled with PM7 semi-empirical method in Gaussian16[46]. To be noted, to save computational costs, not all sizes are sampled. According to the experimental and theoretical predictions, SA-DMA systems tend to grow with the similar molecules number within cluster[5], so for relatively larger cluster in the box, clusters with equal number of acid and base molecules or one more or less than the other are included. Mostly 50000 structures are sampled and then selected by Farthest Point Sampling (FPS) based on Many-body Tensor Representation (MBTR) descriptor[47] for further DFT (wB97XD/6-31++G(d,p)) force calculations to be labelled. Systematic benchmark[48] for aerosol nucleation clusters proves the good balance between accuracy and cost for wB97XD/6-31++G(d,p). The detailed MetaMD sampling and subsequent DFT calculations procedures are listed in Extended Data Table 3.

**Active learning.** Based on the initial data set prepared by MetaMD, active learning or on-the-fly strategy[49] is utilized. The MetaMD sampling subset after screening is the initial data set to kick off the active learning iterations. In each iteration, firstly, 400 thousands steps training with different seeds is conducted to generate four DNN models. Then, NVT ensemble MD simulation is made based on trained DNN models. During the MD, four DNN models are utilized to pinpoint the candidate cluster whose error indicator satisfies the threshold range. The error indicator is the maximal standard deviation of the atomic force predicted by the model ensemble. The threshold ranges from 0.35 eV/Å to 0.50 eV/Å, indicating that those whose error indicator is below 0.35 eV/Å are regarded as accurate and those whose error indicator is above 0.50 eV/Å are regarded as physically unreasonable. Finally, the energies and forces of candidate clusters are calculated through wB97XD/6-31++G(d,p) in Gaussian16[46] package and then put into the data set for training in the next iteration. To be noted, candidate clusters are carved out from the frame according to the interatomic distance cutoff, 3.5 Å. Different from the similar work conducted for combustion reactions[50], here the cluster is obtained as long as the shortest interatomic distance between molecules is short than cutoff value in order to keep integrity of molecular cluster. The detailed iteration processes are listed in Extended Data Table 4.

**DNN model.** The smooth version of deep potential[51,52] model is conducted here. In deep potential, the potential energy of molecular cluster is a sum of "atomic energies" $E = \sum_i E_i$, where $E_i$ is determined by the local environment of atom $i$ within a cutoff radius. The model includes two networks: the embedding network and the fitting network. The embedding network is of size (25, 50, 100) and the fitting network is of size (240, 240, 240). The fitting network uses ResNet architecture[53]. The cutoff radius was set to 6.0 Å and the descriptors decay smoothly from 5.8 to 6.0 Å. The learning rate starts at $1.0 \times 10^{-3}$ and exponentially decays every 2,000 steps in 400,000 training steps in each active learning iteration and exponentially decays every 20,000 steps in 4,000,000 training steps based on the final data set. The loss function is defined as a sum of different mean square errors of the DNN predictions for energy and force.

**Molecular dynamics.** Molecular cluster collision and evaporation simulations are conducted under NVT ensemble through LAMMPS[54] package with 10 SA molecules and 10 DMA molecules initially randomly placing in the cubic box with length of 100 Å. The random positions are given by the aid of packmol[55] package with the stable monomer of SA and DMS as the input. The temperature is kept at 278 K for 17 cases and 300 K for 16 cases respectively. In each case, MD is made firstly for 100 ps with the COM for each molecule being fixed for equilibration and then 1 ns for production. The positions under each frame within production stage are recorded every 10 fs and the cluster index with the interatomic cutoff distance of 3.0 Å is also recorded. The snapshot in Fig. 3a is plotted with VMD[56] while the structure in Fig. 3b is plotted with Chemcraft[57]. The structural clustering analysis in Fig. 3d is conducted using ASAP[58]. The proton transfer distance threshold between O and H is set to 1.23 Å. Collison rate constants are derived according to the Poisson distribution feature of the reactive (collision or evaporation) events[40] using the ChemTraYzer software package[59]. Collision and evaporation events are identified when duration is more than 100 fs[22]. The Poisson-based collision rate constant $k$ can be calculated according to

$$k = \frac{\sum_j N_j}{V \sum_j (\sum_i^M C_i \Delta t_i)_j} \quad (2)$$

Here $N_j$ is the collision events number in case $j$, $V$ is the MD box volume (1000 $nm^3$), $i$ is the subsimulation number in case $j$ separated by reactive events, $C$ is the product of reactants concentration, $\Delta t$ is the interval between reactive events. The detailed derivation for rate constants and confidence interval of Poisson-based reaction events can be found in literature[40].

**Cluster kinetics.** The molecular cluster kinetics simulations are made by our home-built python version[60] of Atmospheric Cluster Dynamics Code (ACDC)[15] solving the ordinary differential equations. The collision rate constants are partially replaced by the MD observed collision events derived constants and the remains are calculated by hard sphere collision model. The evaporation rates are calculated assuming detailed balance based on quantum chemical thermodynamics[20] from literature[61]. The condensation sink (CS) is set to be 2.6 × $10^{-3}$ $s^{-1}$ and 2.7 × $10^{-2}$ $s^{-1}$ to mimic condensation under clean[62] and polluted[63] environment respectively. In ACDC, the formation rate simulation can be calculated by

$$J = \sum_{i=0}^{4}\sum_{j=0}^{4}\sum_{k=0}^{4}\sum_{l=0}^{4} \beta_{ik,jl} c_{ik} c_{jl} \ \ (i + j \ >= 4 \ and \ k + l \ > 4) \quad (3)$$

Here $i$ and $j$ refer to the number of SA molecules in the first and second cluster, $k$ and $l$ refer to the number of DMA molecules. The time evolution of cluster concentration $c_i$ can be obtained by solving the birth and death equations given by

$$\frac{d_{c_i}}{dt} = \frac{1}{2}\sum_{j<i} \beta_{j,(i-j)} c_j c_{i-j} + \sum_j \gamma_{(i+j)\to i} c_{(i+j)} - \sum_j \beta_{i,j} c_i c_j - \frac{1}{2}\sum_{j<i} \gamma_{i\to j} c_i + CS \quad (4)$$

Here $CS$ represents condensation sink. $\beta_{i,j}$ represents the collision rate constants which obtained from hard sphere collision model and is calculated by

$$\beta_{i,j} = \left(\frac{3}{4\pi}\right)^{1/6} \left(\frac{6k_bT}{m_i} + \frac{6k_bT}{m_j}\right)^{1/2} \left(V_i^{1/3} + V_j^{1/3}\right)^2 \quad (5)$$

Here $T$ represents the temperature, $k_b$ represents the Boltzmann constant, and $m_i$ and $V_i$ represent the mass and volume of cluster $i$, respectively. The evaporation coefficient $\gamma_{(i+j)\to i}$ is calculated by

$$\gamma_{(i+j)\to i,j} = \beta_{i,j} \frac{c_i^e c_j^e}{c_{i+j}^e} = \beta_{i,j} c_{\mathrm{ref}} \exp\left(\frac{\Delta G_{i+j} - \Delta G_i - \Delta G_j}{k_b T}\right) \quad (6)$$

**Acknowledgements**

We thank Linfeng Zhang, Jinzhe Zeng and other deep potential community contributors for the diligent support of DPMD and DP-GEN. We thank Linfeng Zhang for very helpful comments on the manuscript. We thank John E. Herr for the help with metadynamics test and analysis. We thank Roope Halonen and Bernhard Reischl for sharing with us the input files to reproduce sulfuric acid monomers collisions simulation. We acknowledge the support of GPU cluster built by MCC Lab of Information Science and Technology Institution, USTC. This work was supported by the National Science Fund for Distinguished Young Scholars (grant no. 41725019), the National Key Research and Development program (grant no. 2016YFC0202203) and the National Natural Science Foundation of China (Grant No. 41877305 and 41605099).

**Author Contributions**

S. J. proposed the framework and conduct passive learning, active learning, DNN training as well as the molecular dynamics and analysis. Y. R. L, T. H, Y. J .F, C. Y. W. and Z. Q. W. helped wrote and edited the paper. W. H. instructed the design of the framework. All authors commented on the manuscript.

**Competing interests**

No potential conflicts of interest exist for any of the listed authors.

**Data and materials availability**

All data related to this study can be obtained from the corresponding author via email.

**Extended Data Table 1 | Root mean squared error (RMSE) of different method performance for describing the dimer detachment (in kcal/mol).**

| | $(SA)_2$ | $(DMA)_2$ | $(SA)_1(DMA)_1$ |
|---|---|---|---|
| PM7 | 1.74 | 1.51 | 14.92 |
| DNN | 1.57 | 1.32 | 0.78 |
| OPLS | 8.73 | Null | Null |

**Extended Data Table 2 | Collision rate constants ($cm^3$ $s^{-1}$) from MD and hard sphere model at 278 K and 300 K.** Those constants' uncertain is below 2.0. k, $k_{low}$ and $k_{up}$ represent the average, upper and lower rate constants from MD while $k_{gas}$ represent those from hard sphere model.

a) T=278 K:

| reaction | k | $k_{low}$ | $k_{up}$ | $k/k_{low}$ | $k_{up}/k$ | $k_{gas}$ | $k/k_{gas}$ |
|---|---|---|---|---|---|---|---|
| $(DMA)_1$+$(DMA)_1$->$(DMA)_2$ | 2.88e-10 | 2.15e-10 | 3.75e-10 | 1.34 | 1.30 | 5.68e-10 | 0.51 |
| $(SA)_1$+$(SA)_1$->$(SA)_2$ | 3.10e-10 | 2.17e-10 | 4.26e-10 | 1.43 | 1.38 | 3.34e-10 | 0.93 |
| $(DMA)_1$+$(SA)_1$->$(SA)_1(DMA)_1$ | 7.39e-10 | 6.18e-10 | 8.76e-10 | 1.20 | 1.19 | 4.53e-10 | 1.63 |
| $(SA)_1$+$(SA)_1(DMA)_1$->$(SA)_2(DMA)_1$ | 8.59e-10 | 5.02e-10 | 1.35e-09 | 1.71 | 1.58 | 4.08e-10 | 2.10 |
| $(DMA)_1$+$(DMA)_2$->$(DMA)_3$ | 2.05e-09 | 1.24e-09 | 3.16e-09 | 1.65 | 1.54 | 6.28e-10 | 3.26 |
| $(SA)_1$+$(SA)_2(DMA)_1$->$(SA)_3(DMA)_1$ | 2.54e-09 | 1.35e-09 | 4.27e-09 | 1.88 | 1.68 | 4.31e-10 | 5.90 |
| $(SA)_1$+$(SA)_2$->$(SA)_3$ | 1.64e-08 | 1.30e-08 | 2.04e-08 | 1.26 | 1.24 | 3.69e-10 | 44.39 |

b) T=300 K:

| reaction | k | $k_{low}$ | $k_{up}$ | $k/k_{low}$ | $k_{up}/k$ | $k_{gas}$ | $k/k_{gas}$ |
|---|---|---|---|---|---|---|---|
| $(DMA)_1$+$(DMA)_1$->$(DMA)_2$ | 4.04e-10 | 3.07e-10 | 5.20e-10 | 1.32 | 1.29 | 5.90e-10 | 0.68 |
| $(DMA)_1$+$(SA)_1(DMA)_1$->$(SA)_1(DMA)_2$ | 5.82e-10 | 3.71e-10 | 8.61e-10 | 1.57 | 1.48 | 5.89e-10 | 0.99 |
| $(DMA)_1$+$(SA)_1$->$(SA)_1(DMA)_1$ | 6.81e-10 | 5.58e-10 | 8.20e-10 | 1.22 | 1.20 | 4.70e-10 | 1.45 |
| $(SA)_1$+$(SA)_1$->$(SA)_2$ | 5.19e-10 | 3.81e-10 | 6.86e-10 | 1.36 | 1.32 | 3.47e-10 | 1.50 |
| $(SA)_1$+$(SA)_1(DMA)_1$->$(SA)_2(DMA)_1$ | 6.43e-10 | 3.95e-10 | 9.79e-10 | 1.63 | 1.52 | 4.24e-10 | 1.52 |
| $(SA)_1$+$(SA)_2(DMA)_1$->$(SA)_3(DMA)_1$ | 1.59e-09 | 8.21e-10 | 2.73e-09 | 1.94 | 1.72 | 4.47e-10 | 3.55 |
| $(DMA)_1$+$(SA)_2$->$(SA)_2(DMA)_1$ | 2.20e-09 | 1.37e-09 | 3.31e-09 | 1.61 | 1.51 | 5.46e-10 | 4.03 |
| $(DMA)_1$+$(DMA)_2$->$(DMA)_3$ | 3.42e-09 | 2.15e-09 | 5.10e-09 | 1.59 | 1.49 | 6.52e-10 | 5.24 |
| $(DMA)_1$+$(SA)_3(DMA)_1$->$(SA)_3(DMA)_2$ | 1.32e-08 | 7.22e-09 | 1.45e-08 | 1.82 | 1.10 | 6.98e-10 | 18.91 |
| $(SA)_1$+$(SA)_2$->$(SA)_3$ | 8.53e-09 | 5.70e-09 | 1.22e-08 | 1.50 | 1.43 | 3.84e-10 | 22.23 |
| $(SA)_1$+$(SA)_4(DMA)_1$->$(SA)_5(DMA)_1$ | 2.12e-08 | 1.09e-08 | 3.64e-08 | 1.94 | 1.72 | 5.08e-10 | 41.70 |

**Extended Data Table 3 | Passive learning and active learning sampling structure number for the corresponding cluster composition.**

| Cluster composition | Passive learning | | Active learning |
|---|---|---|---|
| | Metadynamics sampling | DFT calculations | DFT calculations |
| $(DMA)_1$ | 99999 | 69678 | 2 |
| $(DMA)_2$ | 50000 | 15536 | 22 |
| $(DMA)_3$ | 100000 | 4998 | 2 |
| $(DMA)_4$ | 0 | 0 | 13 |
| $(SA)_1$ | 257973 | 58872 | 2 |
| $(SA)_1(DMA)_1$ | 120000 | 25171 | 84 |
| $(SA)_1(DMA)_2$ | 50000 | 10532 | 447 |
| $(SA)_1(DMA)_3$ | 0 | 0 | 54 |
| $(SA)_1(DMA)_4$ | 0 | 0 | 57 |
| $(SA)_2$ | 144320 | 4610 | 51 |
| $(SA)_2(DMA)_1$ | 49999 | 16377 | 63 |
| $(SA)_2(DMA)_2$ | 50000 | 4339 | 106 |
| $(SA)_2(DMA)_3$ | 50000 | 4998 | 13 |
| $(SA)_2(DMA)_4$ | 0 | 0 | 201 |
| $(SA)_2(DMA)_5$ | 0 | 0 | 14 |
| $(SA)_3$ | 50303 | 5028 | 24 |
| $(SA)_3(DMA)_1$ | 0 | 0 | 3 |
| $(SA)_3(DMA)_2$ | 49863 | 4801 | 62 |
| $(SA)_3(DMA)_3$ | 49998 | 4992 | 51 |
| $(SA)_3(DMA)_4$ | 49997 | 4987 | 12 |
| $(SA)_4$ | 50000 | 4014 | 0 |
| $(SA)_4(DMA)_3$ | 49997 | 4991 | 146 |
| $(SA)_4(DMA)_4$ | 49867 | 2315 | 0 |
| $(SA)_4(DMA)_5$ | 50293 | 200 | 17 |
| $(SA)_5(DMA)_4$ | 49999 | 200 | 0 |
| $(SA)_5(DMA)_5$ | 50000 | 200 | 109 |
| $(SA)_5(DMA)_6$ | 49999 | 191 | 0 |
| Total | 1522607 | 247030 | 1555 |
| Percentage of total DFT structure number | Null | 99.37% | 0.63% |

**Extended Data Table 4 | Molecular dynamics parameters in active learning iterations.**

| Iteration number | Initial molecular composition[a] | box length | length(ps) | structures number[b] |
|---|---|---|---|---|
| 0 | (4, 4) | 40 | 2 | 13 |
| 1 | | | 2 | 27 |
| 2 | | | 10 | 36 |
| 3 | | | 10 | 103 |
| 4 | | | 20 | 89 |
| 5 | | | 50 | 172 |
| 6 | (5, 5) | 50 | 100 | 168 |
| 7 | | | 200 | 480 |
| 8 | | | 500 | 467 |

Notes: a. (m, n) represents the number of SA and DMA molecules respectively. b. structures number gives the total cluster structures number derived from each active learning iteration.

**Extended Data Figure 1 | Formation of $(SA)_6(DMA)_6$ in DNN-MD arising from the collision of $(SA)_1(DMA)_1$ and $(SA)_5(DMA)_5$.**

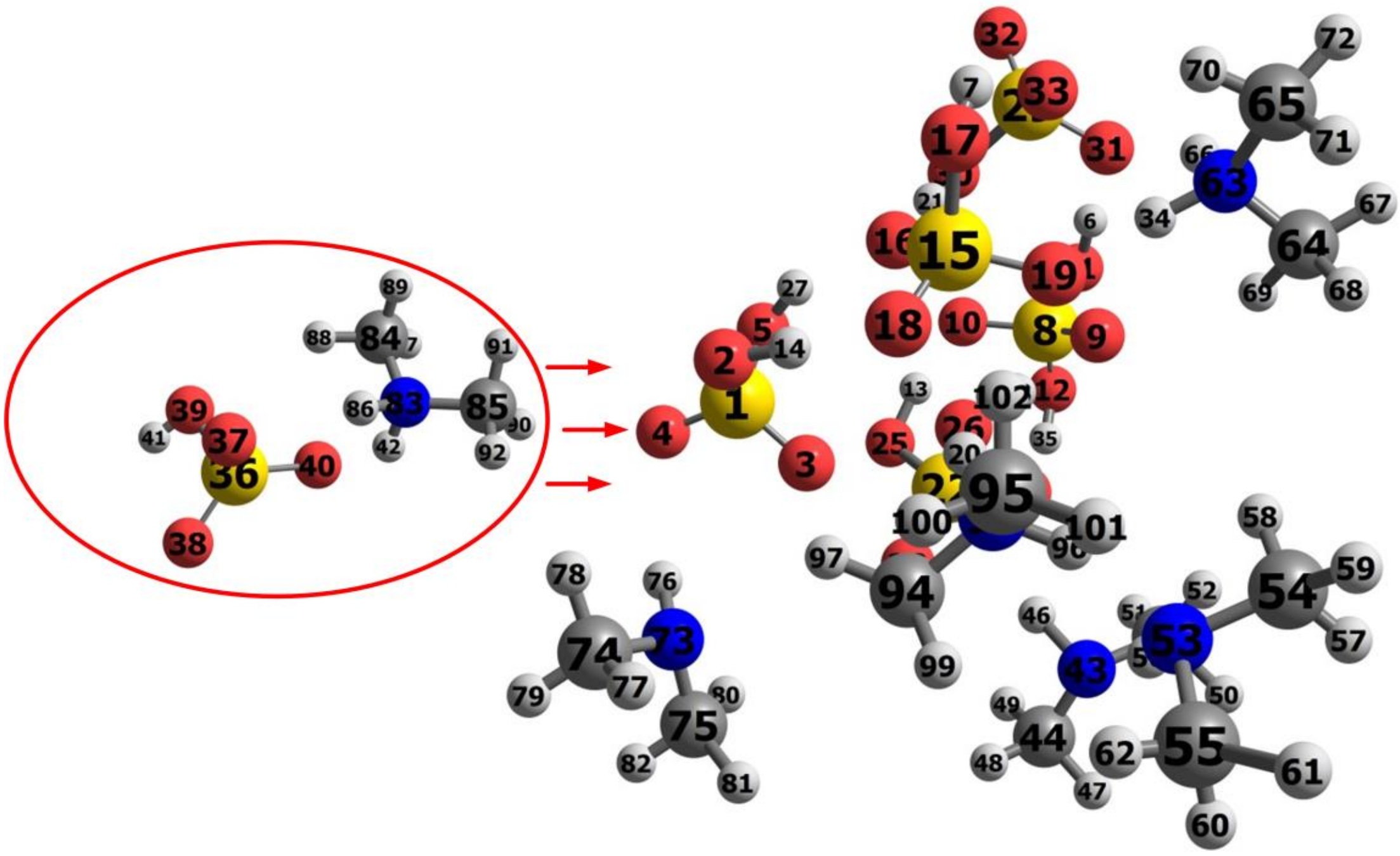

**Extended Data Figure 2 | Cluster shape anisotropy for each composition with uncertain bar at 278 K.**

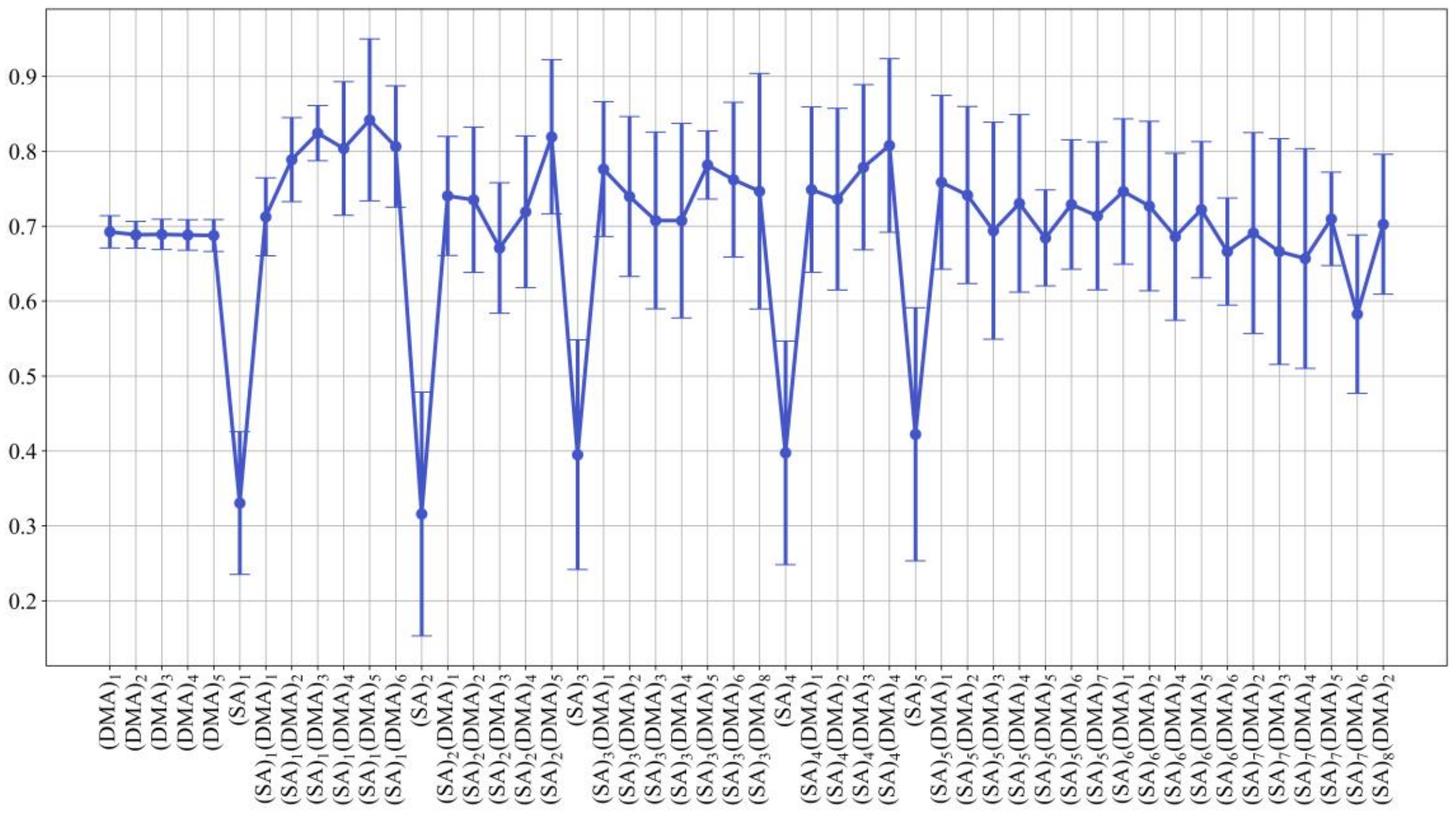

**Extended Data Figure 3 | Cluster shape anisotropy for each composition with uncertain bar at 300 K.**

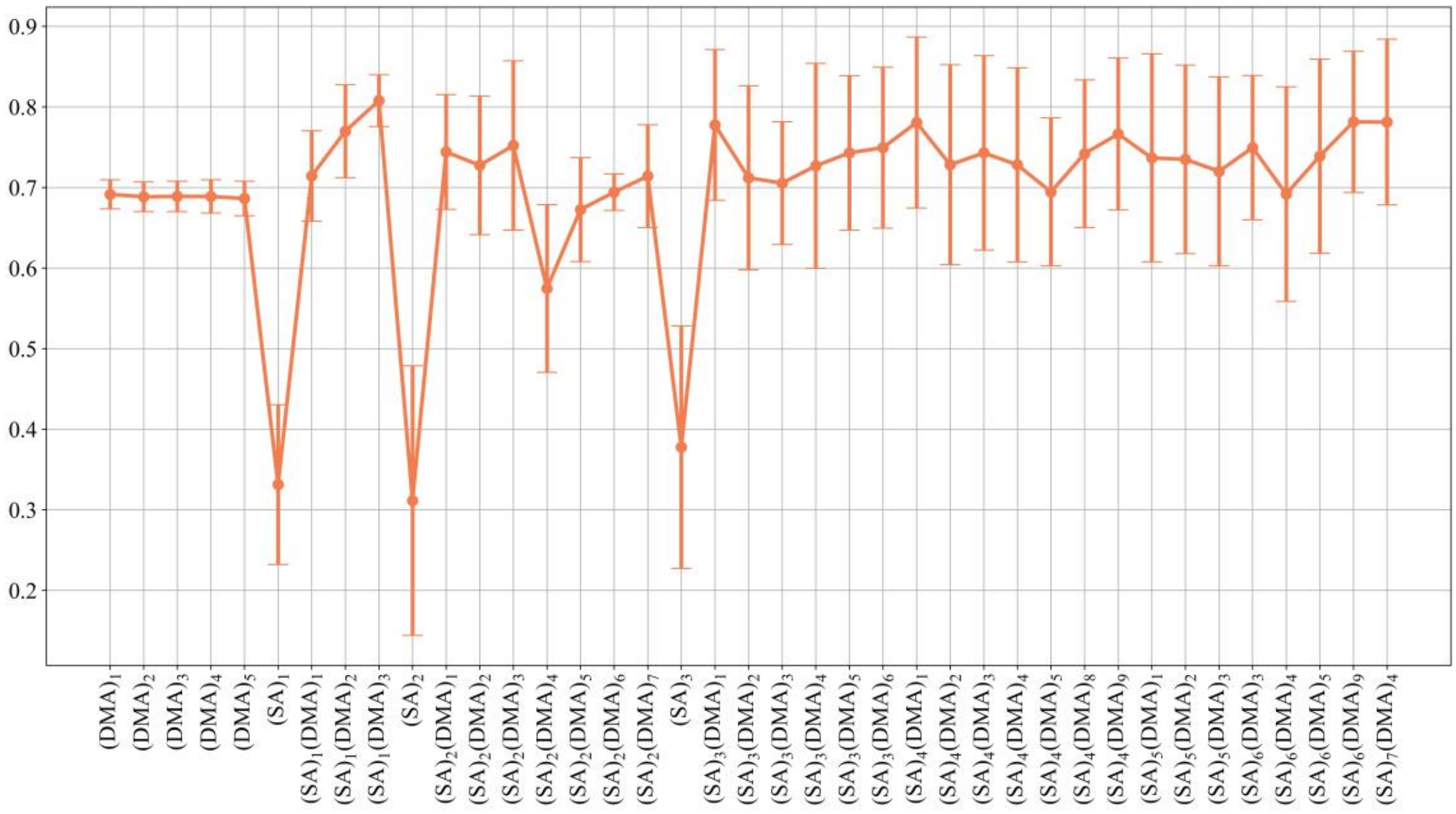

**Extended Data Figure 4 | H and O atom distance for each sulfuric acid in $(SA)_6(DMA)_6$ cluster.** Nine H-O pairs with covalent bonds in sulfuric acid molecules observed in the initial moment of $(SA)_6(DMA)_6$ formation are plotted below.

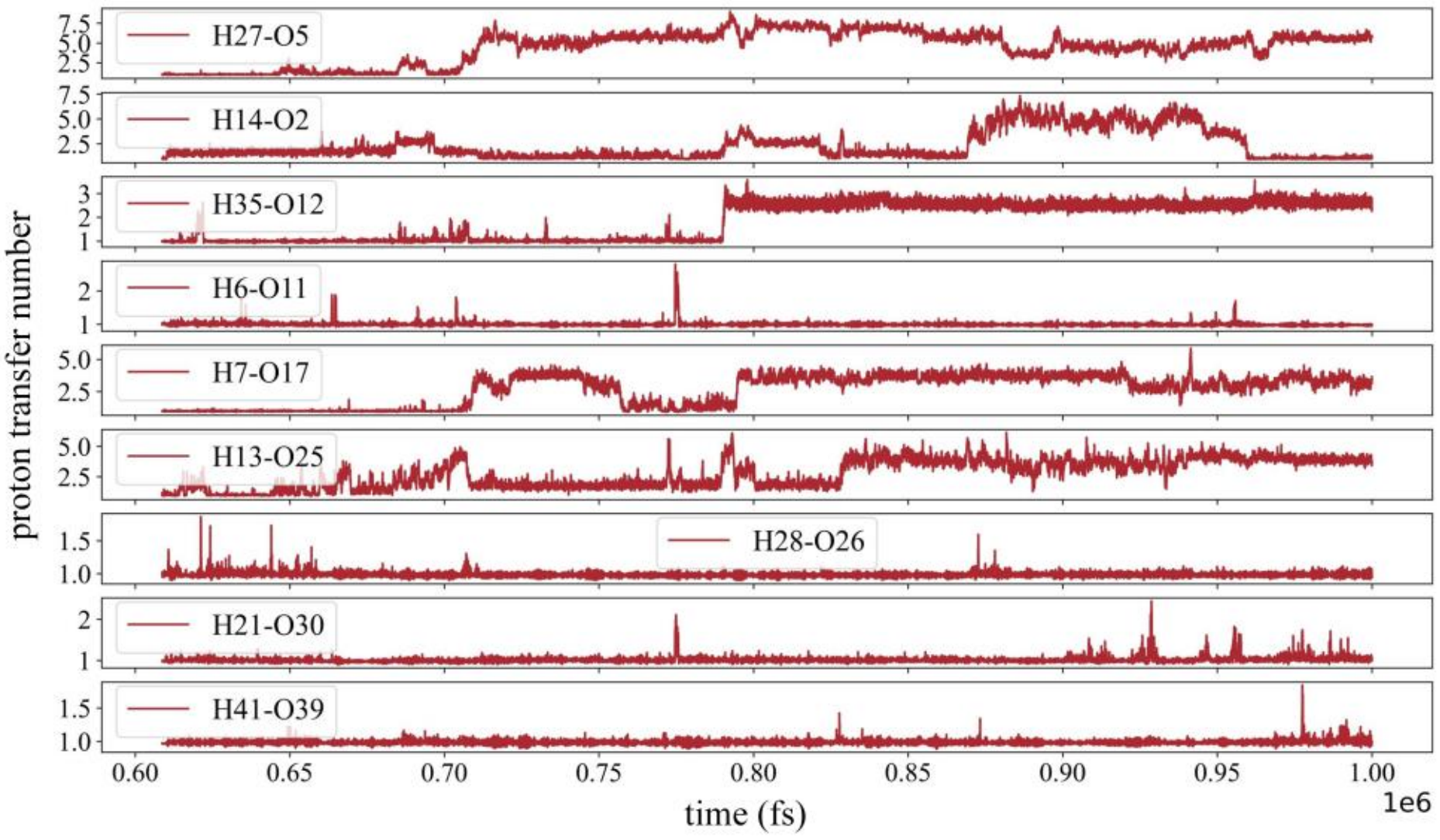

**Extended Data Figure 5 | Time evolution of nine hydrogen nitrogen pairs distance for $(SA)_6(DMA)_6$ cluster.**

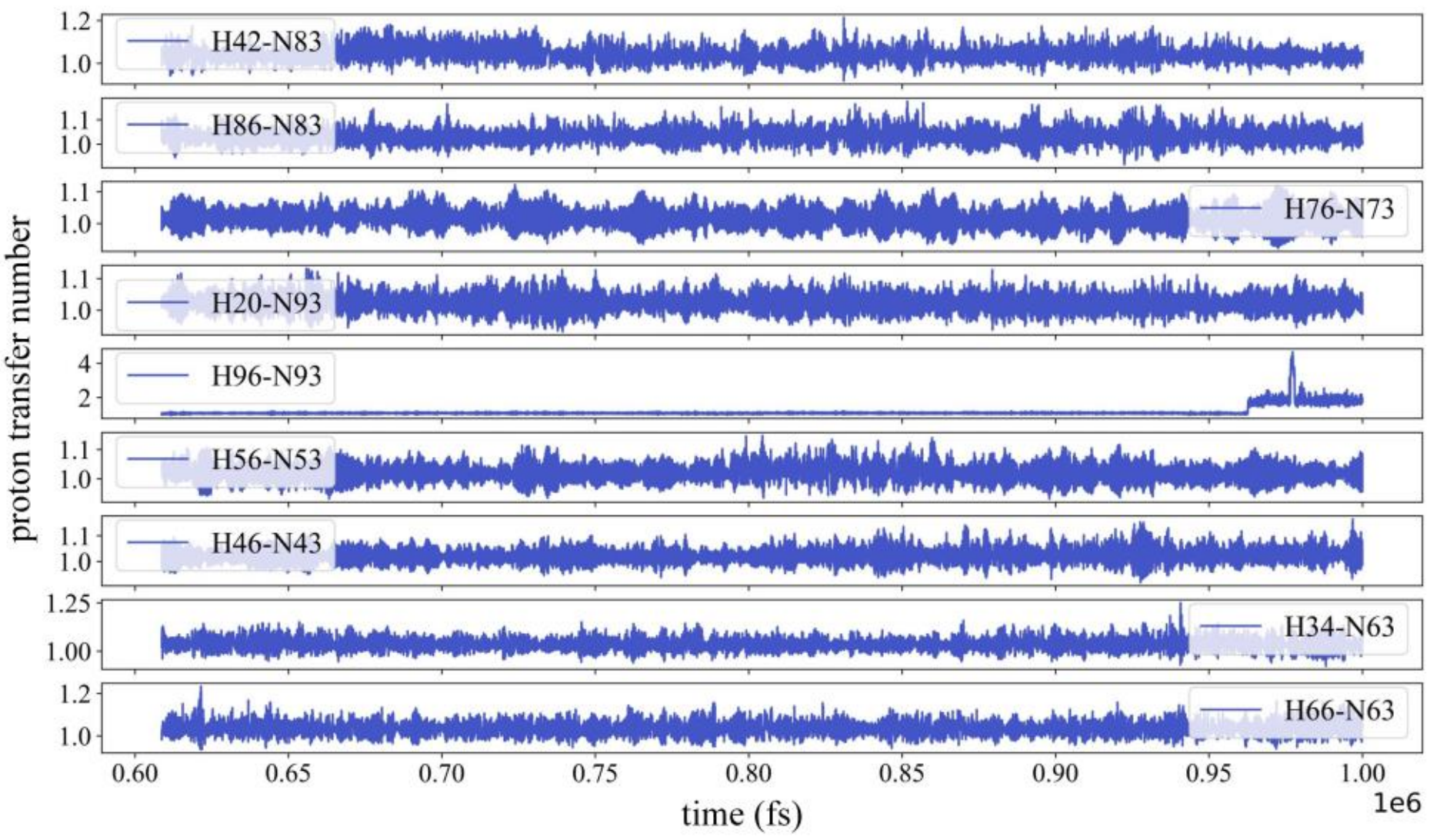

**Extended Data Figure 6 | Time evolution of proton number within each sulfuric acid molecule for $(SA)_6(DMA)_6$ cluster. To be noted, the time evolution is filtered with the proton number duration being more than 100 fs for observing relatively long existence of bonds.**

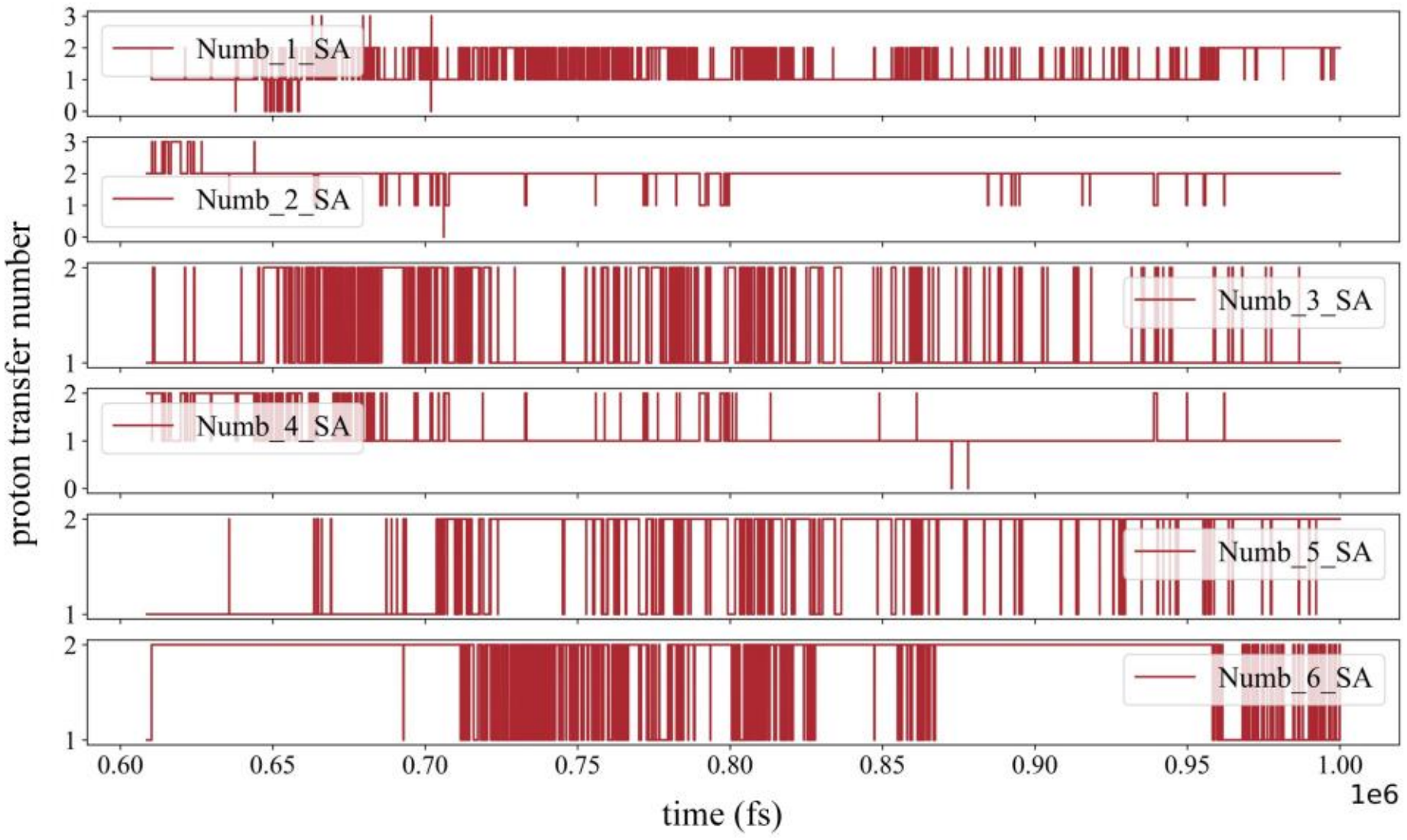

**Extended Data Figure 7 | Radial distribution function (RDF) for clusters with the same number of acid and base molecules inside at 278 K.**

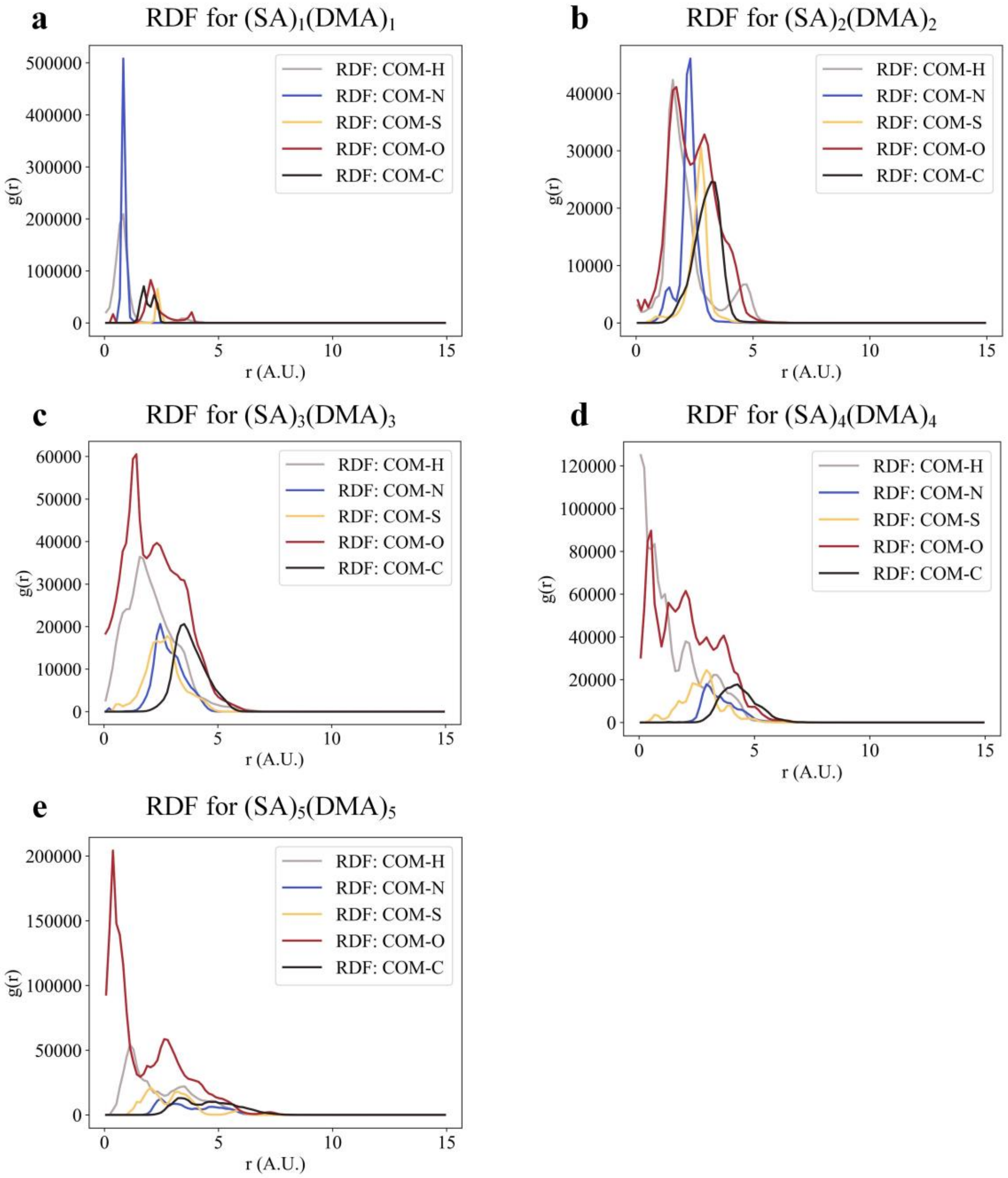

**Extended Data Figure 8 | Cluster shape anisotropy distribution across all structures for cluster $(SA)_m(DMA)_n$ (m=0~4, n~0~4) at 278 K.**

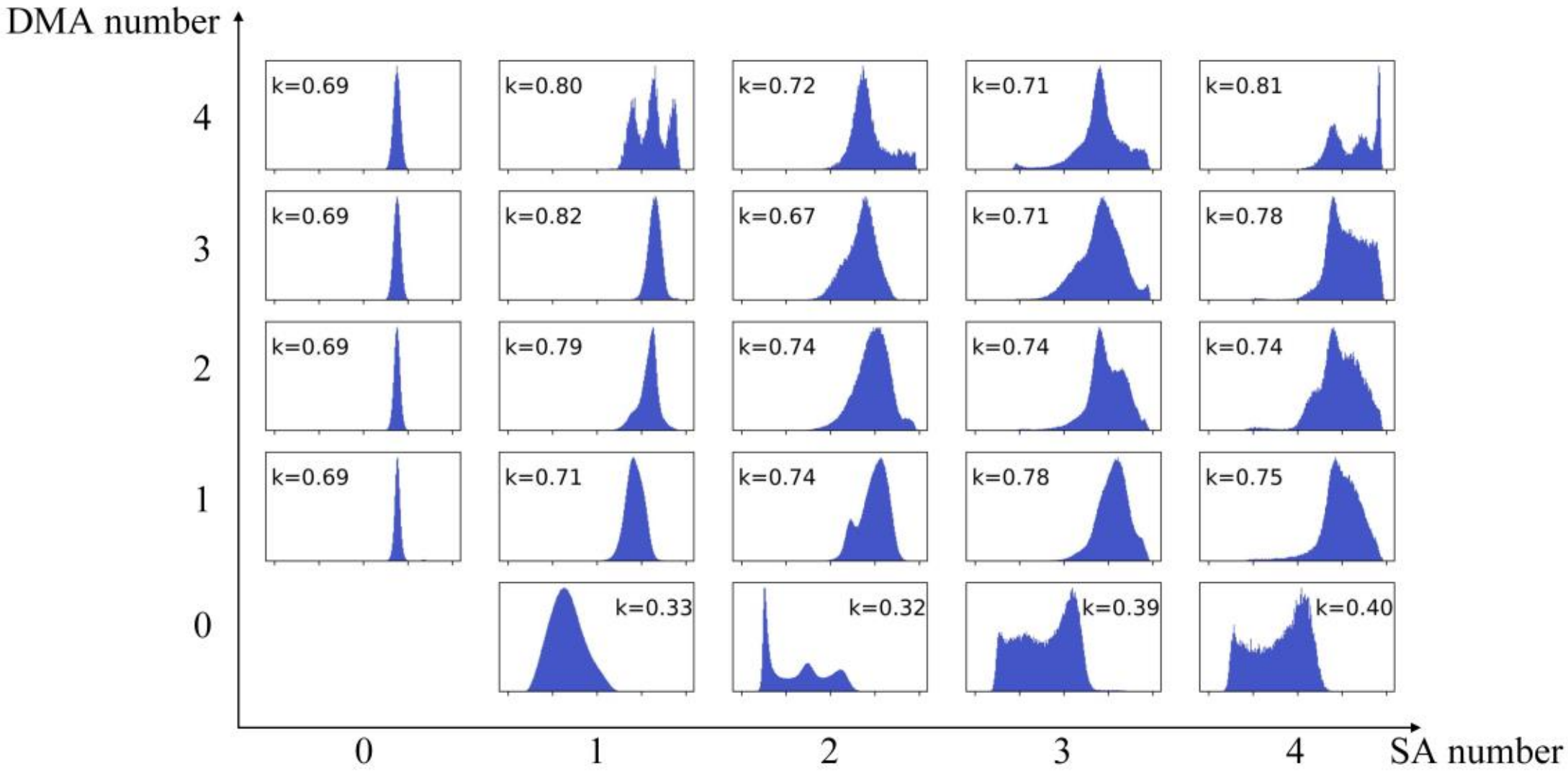